\documentclass[aps,prl,twocolumn,superscriptaddress, floatfix]{revtex4-2}

\usepackage{gensymb}  
\usepackage{graphicx}  
\usepackage{hyperref}
\usepackage[table]{xcolor}

\hypersetup{colorlinks=true,urlcolor= blue,citecolor=blue,linkcolor= blue}

\usepackage{enumitem}
\setlist{noitemsep,leftmargin=*,topsep=0pt,parsep=0pt}
\usepackage{chemmacros}
\usepackage{upgreek}
\usepackage{placeins} 

\begin{document}

\title{Interplay of Defects and the Charge Density Wave State in Hf-Doped ZrTe$_{3}$}



\author{Ghilles Ainouche}
\author{Resmi Sudheer}
\author{Susree Mohapatra}
\author{Boning Yu}
\author{Muhammad Suhayb Malik}
\affiliation{Department of Physics, Clark University, Worcester, Massachusetts 01610, USA}
\author{Yu Liu}
\affiliation{Condensed Matter Physics and Materials Science Division, Brookhaven National Laboratory, Upton, New York 11973, USA}
\affiliation{Current address: Center for Correlated Matter and School of Physics, Zhejiang University, Hangzhou 310058, China}
\author{Cedomir Petrovic}
\affiliation{Condensed Matter Physics and Materials Science Division, Brookhaven National Laboratory, Upton, New York 11973, USA}
\affiliation{Current address: Shanghai Key Laboratory of Material Frontiers Research in Extreme Environments (MFree), Shanghai Advanced Research in Physical Sciences (SHARPS), Shanghai 20120, China}
\author{Abhilash Ravikumar}
\affiliation{Department of Electronics and Communication Engineering, Amrita School of Engineering, Amrita Vishwa Vidyapeetham, Bengaluru, India}
\author{Michael C. Boyer}
\email{To whom correspondence should be addressed. mboyer@clarku.edu}
\affiliation{Department of Physics, Clark University, Worcester, Massachusetts 01610, USA}

\date{\today}

\begin{abstract}
We carry out temperature-dependent scanning tunneling microscopy (STM) studies of the charge density wave (CDW) compound ZrTe$_3$ which is intentionally doped with Hf. Previous bulk studies tie Hf doping to an enhancement of the CDW transition temperature (T$_{CDW}$). In our work, by combining STM measurements with density functional theory (DFT) calculations, we observe and identify multiple defects in Zr$_{0.95}$Hf$_{0.05}$Te$_3$. Surprisingly, instead of finding clear structural or electronic signatures associated with Hf dopants, we determine the origin of the observed defects are consistent with Te and Zr vacancies. Further, our temperature dependent STM measurements allow us to examine CDW pinning to both types of observed defects below and above T$_{CDW}$.
\end{abstract}

\maketitle



Manipulating material properties through the purposeful (or accidental) introduction of dopants/defects has brought insights into the fundamental physics hosted within as well as an understanding of how material properties can be optimized for use in applications.\cite{Mishra2024, Zhussupbekov2021, Hart2024, Alloul2009, Lin2023, Hossen2024} Scanning tunneling microscopy (STM) has been a useful tool to study the real-space atomic-scale effects that defects have on material quantum states.\cite{Garnica2022, Hudson2001, Pan2000, Hanneken2016, Kim2019, Mottas2019} Here we combine STM measurements with density functional theory (DFT) calculations to study the charge density wave compound ZrTe$_3$ doped with Hf.

ZrTe$_3$ is a member of the transition metal tri-chalcogenides (MX$_3$ where M = transition metal ion and X = chalcogen anion) and hosts a unidirectional incommensurate charge density wave (CDW) state below T$_{CDW}$ $\sim$ 63 K which coexists with a superconducting state below T$_{SC}$ $\sim$ 2 K.\cite{Yamaya2012, Zhu2016} The CDW is believed to originate from Fermi surface nesting with a CDW wavevector of \textit{q} = 0.071 \textit{a*} + 0.336 \textit{c*} determined by electron diffraction measurements \cite{Eaglesham1984}, consistent with nesting of quasi-one dimensional Fermi surface sheets established by the 5p$_x$ band associated Te-Te chains along the crystallographic \textit{a}-axis \cite{Yokoya2005, Hoesch2009}. However, it is believed that momentum-dependent electron-phonon coupling, in addition to Fermi surface nesting, also plays an important role in the CDW.\cite{Hu2015}

Previous studies of ZrTe$_3$ show that T$_{CDW}$ can be manipulated by the introduction of intentionally doped defects. Se substituting for Te in ZrTe$_{3-x}$Se$_x$ leads to a decrease in T$_{CDW}$ and increases in T$_{SC}$ with a complete suppression of the CDW state for x $<$ 0.03 and a T$_{SC}$ maximum of 4.4 K.\cite{Zhu2016} The Se dopants suppress the CDW state by introducing disorder into the \textit{a}-axis Te-Te chains which form the parallel Fermi surface sheets.\cite{Liu2022} On the other hand, substituting Hf for Zr increases T$_{CDW}$ with T$_{CDW}$ $\sim$ 72 K for Zr$_{0.95}$Hf$_{0.05}$Te$_3$; Hf dopants preserve the \textit{a}-axis Te-Te chains while enhancing electron-phonon coupling.\cite{Liu2022}

\begin{figure}[!hb]
\includegraphics[clip=true,width=\columnwidth]{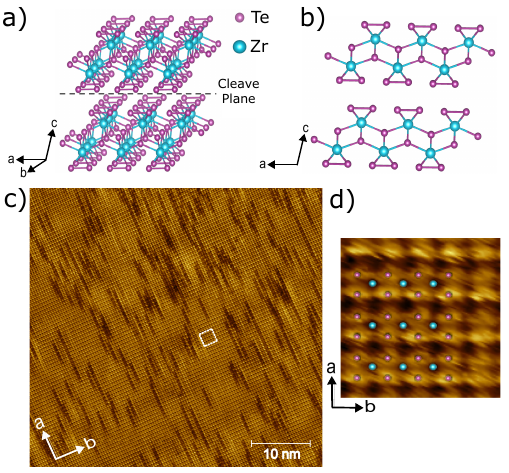}
\caption{Crystal structure and topography of ZrTe$_3$: (a) Crystal structure showing the layered nature of ZrTe$_3$, highlighting the cleavage plane. (b) Crystal structure in the \textit{a-c} plane, depicting the trigonal chains oriented along the \textit{b}-axis. Figures a) and b) were created using Vesta.\cite{Momma2011} (c) 50 nm x 45 nm STM topography taken at 9 K, showing atomic resolution, defects, and the unidirectional CDW along the \textit{a}-axis (I = 100 pA, V$_{Sample}$ = -420 mV). (d) Zoom of region away from defects, with an overlaid structural model of the \textit{a-c} plane showing the surface Te ions imaged directly by STM as well as the neighboring Zr ions below.}
\label{fig:Fig1}
\end{figure}

\begin{figure}[t]
\includegraphics[clip=true,width=\columnwidth]{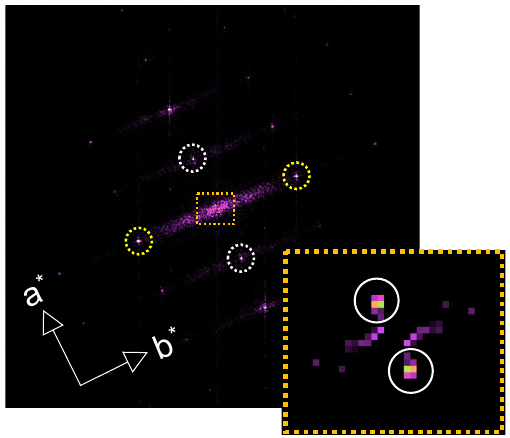}
\caption{FFT of topographic image from Figure \ref{fig:Fig1}c. Peaks associated with atomic lattice periodicities are circled in dotted white and yellow. The orange box is a zoom of the center of the FFT showing \textit{q}$_{CDW}$ (circled in solid white).}
\label{fig:Fig2}
\end{figure}

Such tunability of ZrTe$_3$ properties with doping make the compound of significant interest in understanding how the hosted quantum states are affected at the atomic-scale and for their potential use in applications. In this work we use STM to study Zr$_{0.95}$Hf$_{0.05}$Te$_3$ from 9 K (well-below T$_{CDW}$) to 92 K (above T$_{CDW}$). In our measurements, we identify multiple defects which we conclude are due to Te and Zr vacancies, not Hf dopants. We study the temperature evolution of the interplay of the CDW state with defects through T$_{CDW}$.

ZrTe$_3$ is a layered compound comprised of ZrTe$_6$ trigonal prism chains extending along the \textit{b}-crystal axis (Figure \ref{fig:Fig1}a,b).\cite{Felser1998} Neighboring layers are weakly bound together via Van der Waals forces which lead to natural cleave planes. Consequently, STM measurements directly probe the surface Te ions. Figure \ref{fig:Fig1}c shows a topographic image of ZrTe$_3$ acquired at 9 K. Figure \ref{fig:Fig1}d is a zoomed image of the crystal lattice which illustrates an apparent inequivalence between the surface Te ions Te(2) and Te(3). Indeed, in the crystal structure, the Te(2) ions are very slightly elevated along the \textit{c}-axis relative to the Te(3) ions.\cite{Stowe1998} This leads to the Te(2) ions appearing slightly brighter than the Te(3) ions in STM topographic images.

\begin{figure}[h]
\includegraphics[clip=true,width=\columnwidth]{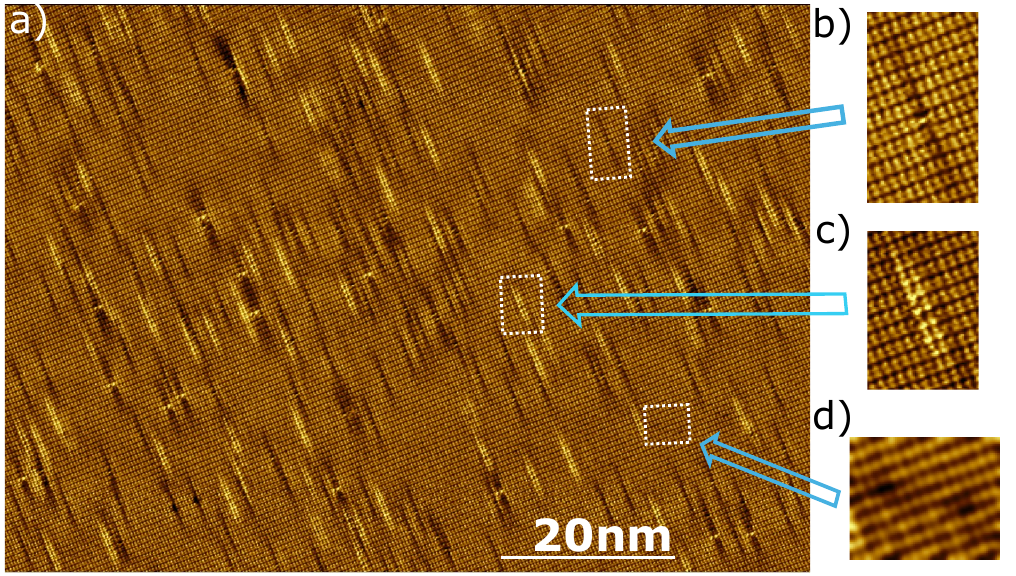}
\caption{(a) 92 nm x 64 nm topography taken at 92 K (I = 100 pA, V$_{Sample}$ = -70 mV) illustrating numerous defects throughout. Two main types of defects are seen in images at all temperatures in this study: EDDs (b) and EBDs (c). (d) An additional type of observed defect: Region with two dark, localized defects which are not EDDs. Gaussian smoothing (5 px) is needed to more-easily identify these types of defects within the image.}
\label{fig:Fig3}
\end{figure}

In addition to the surface Te ions, broad stripes are evident across the sample surface. The stripes are a visualization of the CDW erected along the crystallographic \textit{a}-axis due to connected 5p$_x$ orbitals of the Te ions along this axis. Analysis of the Fast Fourier Transform (FFT) of the topography (Figure \ref{fig:Fig2}) shows the CDW wavevector to be \textit{q$_{CDW}$} =  0.07\textit{a}* which is in agreement with electron \cite{Eaglesham1984} and x-ray \cite{Hoesch2016} diffraction measurements. As STM probes the \textit{a-b} plane of ZrTe$_3$, we are unable to detect the \textit{c}* component of the CDW accessible to diffraction techniques. Notable also in the topographic images, and the focus of this work, are extended defects along the \textit{a}-axis.

In our measurements, at negative biases ranging from at least -50 mV to -400 mV \cite{SupMat}, we find two main types of defects which are extended in nature and exist over the temperature range of our studies (9 K to 92 K). Figure \ref{fig:Fig3} shows a topographic image acquired at 92 K containing and illustrating these defects, one type appears dark and the other bright.\newline

\begin{figure}[h]
\includegraphics[clip=true,width=\columnwidth]{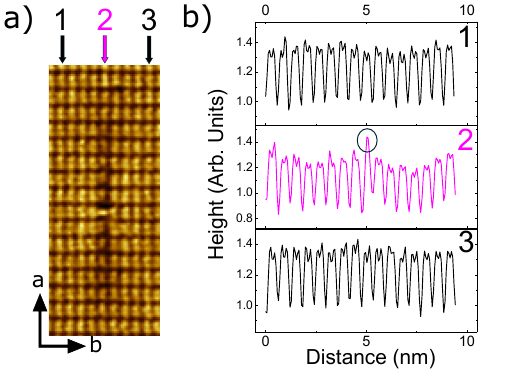}
\caption{a) Topography (10.0 nm x 4.2 nm) around an EDD acquired at 92.6 K (I = 50 pA, V$_{Sample}$ = -50 mV). b) Line cuts taken through (2) and away (1 and 3) from the defect. The linecut through the defect affected region shows a single peak instead of a double peak, indicating a missing Te ion.}
\label{fig:Fig4}
\end{figure}

\noindent{\textit{Extended Dark Defects:}}
An isolated extended dark defect (EDD) is seen in Figure \ref{fig:Fig4}a. Whereas the defect-affected region is extended along the \textit{a}-axis, its effects are limited to one unit cell along the \textit{b}-axis. To pinpoint the structural characteristics of the defect, we take linecuts along the \textit{a}-axis both through and well away from the defect-affected area (Figure \ref{fig:Fig4}b).

Linecuts 1 and 3, taken away from the defect, show a repeated pattern of two smaller peaks appearing upon broader peaks. The broader peak periodicity is due to the repeating unit cell. The two smaller peaks on top correspond to the Te(2) and Te(3). Linecut 2 goes through the defect affected region. In this linecut we observe a slight depression entering the defect region which is not seen in linecuts 1 and 3 explaining the darker appearance of the defect in the topography. This depression emphasizes the extended nature of the defect spanning roughly 14 or 15 unit cells. In addition to this depression, a single peak with high intensity is observed at the center of the defect instead of two. This indicates that the defect is likely a Te vacancy with the remaining Te ion of the Te(2)-Te(3) pair relaxing to a more central location.

A Te vacancy introduces a structural component to the defect. A missing atom introduces a local structural distortion and unlikely explains the extended nature of the defect which preferentially aligns with the \textit{a}-axis. Rather, this extended nature of the defect is likely electronic in nature. Similar to Se-doped ZrTe$_3$ \cite{Liu2022} a Te vacancy will introduce disorder in the \textit{a}-axis Te-Te chains, disrupting the extended 5p$_x$ orbitals running along that direction.

While the linecuts strongly suggest the defect origin as a Te surface vacancy, we turn to DFT calculations for confirmation. DFT calculations were performed using Quantum ESPRESSO using norm-conserving, ultrasoft, and projector-augmented wave (PAW) methods \cite{Blochl1994}. STM simulations were generated using the Tersoff-Hamann approach \cite{Tersoff1983}. The ZrTe$_3$ surface layer was modeled using a 7x3x1 supercell corresponding to dimensions of 1.17 \text{\AA} x 40.6 \text{\AA} in the \textit{a-b} crystal plane. First, the supercell was relaxed, allowing the atomic positions to relax until the forces converged below 0.05 eV/\text{\AA}. The wavefunction energy cutoff was set to 50 Ry (680 eV), and the charge density cutoff to 400 Ry (5442 eV). The Brillouin zone was sampled using a 2x3x3 \textit{k}-point grid for relaxation. Using the same parameters as the relaxation step, self-consistent field calculations were performed to determine the ground-state electronic structure of the system. This was achieved by iteratively solving the Kohn-Sham equations until a converged electron density was obtained, allowing for the extraction of the Kohn-Sham eigenvalues. Next, non-self-consistent field calculations were performed using a finer 6x9x9 \textit{k}-point grid to compute the electronic properties necessary for calculating the local density of states, ultimately allowing for creation of the simulated STM images.

\begin{figure}[t]
\includegraphics[clip=true,width=\columnwidth]{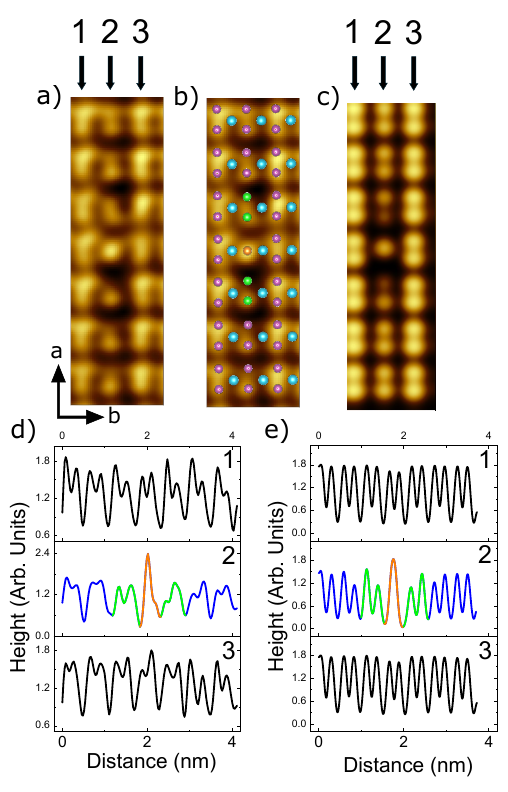}
\caption{Extended dark defect STM image and simulation. a) STM-acquired image (4.06 nm x 1.17 nm) zoomed in on an EDD (I = 50 pA, V$_{Sample}$ = -50 mV). b) Calculated relaxed structure with single Te vacancy overlaid on the STM-acquired image from a). The ions shown in green/red are Te ions color coded to help associate features in the linecuts of d/e with ions in the topographies. c) Simulated -50 mV topography. d) Three \textit{a}-axis linecuts through STM image with the center linecut through the defect region. For clarity of presentation, linecut 2 is presented on a slightly different y-scale. e) Corresponding three \textit{a}-axis linecuts through the simulated image showing general agreement with those through the STM image.}
\label{fig:Fig5}
\end{figure}

Figure \ref{fig:Fig5} shows a) high-resolution STM image zoomed on an EDD, b) an overlay of the relaxed structure highlighting correspondence between the locations of the relaxed atomic positions and ions in the STM image, and c) the simulated STM topography. A comparison of additional linecuts through the STM-acquired topography (Figure \ref{fig:Fig5}d) and simulated topography (Figure \ref{fig:Fig5}e) emphasize the similarity in features. Prominently, at the center of the defect, is a single central peak (highlighted in orange), an indicator of a missing Te ion and a relaxed positioning of the remaining Te ion of the Te(2)-Te(3) pair. In addition, the simulation reproduces the extended single-column depressed/dark nature of the defect along the a-axis compared to the neighboring Te columns. Taken together, the simulation and STM measurements on EDDs strongly support Te vacancies as the origin of these defects.\newline

\begin{figure}[h]
\includegraphics[clip=true,width=\columnwidth]{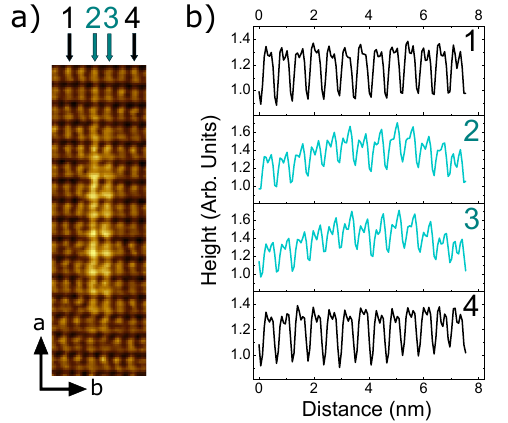}
\caption{Extended bright defect STM image and linecuts. a) Topography (8.0 nm x 2.7 nm) around an EBD acquired at 92.6 K (I = 50 pA, V$_{Sample}$ = -50 mV). b) Line cuts taken through (2, 3) and away (1, 4) from the defect. The linecuts through the defect affected region rise toward the center and have a slight depression at the defect center. For clarity of presentation, linecuts 2 and 3 are presented using a slightly different y-scale.}
\label{fig:Fig6}
\end{figure}

\noindent{\textit{Extended Bright Defects:}}
The second main type of defects we observe in our topographic images of ZrTe$_3$ are extended bright defects (EBDs). Figure \ref{fig:Fig6} shows an isolated EBD. Similar to EDDs, EBDs extend along the \textit{a}-axis, but the spatial extent is across two unit cells along the \textit{b}-axis instead of the one for the EDDs. Linecuts through and away from the EBD illustrate changes within the defect-affected region. First, the linecuts show no evidence for major structural distortions associated with the EBD, pointing toward a sub-surface origin for the defect. Next, the bright appearance of the EBD is seen by the rising nature of the linecuts from either EBD edge toward the defect center. These rising linecuts show that the EBD affects at least 12 unit cells along the \textit{a}-axis. We note a slight depression in the raised regions near the center of the EBD which affects a few unit cells.

Whereas the origin of EDDs could reasonably be deduced from STM measurements alone, here we rely on DFT calculations to determine the origin of the observed EBDs. Given that our samples are intentionally doped with Hf, with Hf substituting for Zr, it is natural to first examine the possibility that the subsurface defects are due to Hf doping.

\begin{figure}[b]
\includegraphics[clip=true,width=\columnwidth]{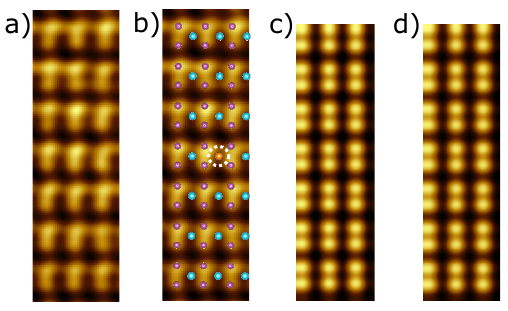}
\caption{Pristine and Hf Doping: a) STM-acquired image (4.06 nm x 1.17 nm) away from any obvious defects (I = 50 pA, V$_{Sample}$ = -50 mV). b) Calculated relaxed structure overlaid on the STM-acquired image from a). The circled orange ion represents a Zr ion location. In the pristine -50 mV simulated topography (c) the orange ion is a Zr ion. In the Hf doped -50 mV simulated topography (d), the orange ion is a Hf ion. In both cases, the relaxed structure overlaying the STM image in b) is identical. Hf doping does not visibly change the STM topography.}
\label{fig:Fig7}
\end{figure}

Figure \ref{fig:Fig7} shows a) high-resolution STM image away from any defects, b) an overlay of the relaxed defect-free structure highlighting correspondence between the locations of the relaxed atomic positions and ions in the STM image, c) a simulated defect-free STM topography, d) a simulated STM topography with a Hf dopant substituting for a subsurface Zr ion. There are two items to note. First, the DFT simulation well-reproduces the STM-acquired topography. Second, the substitution of an Hf dopant leads to no noticeable surface effects. In short, it is unlikely that the EBD originates from Hf dopants. This is not necessarily surprising as Hf dopants were previously noted to preserve the Te(2)-Te(3) chains along the a-axis in ZrTe$_3$ \cite{Liu2022}. This is likely due to Hf being isoelectronic with, and having nearly an identical atomic radius to, Zr.

\begin{figure}[t]
\includegraphics[clip=true,width=\columnwidth]{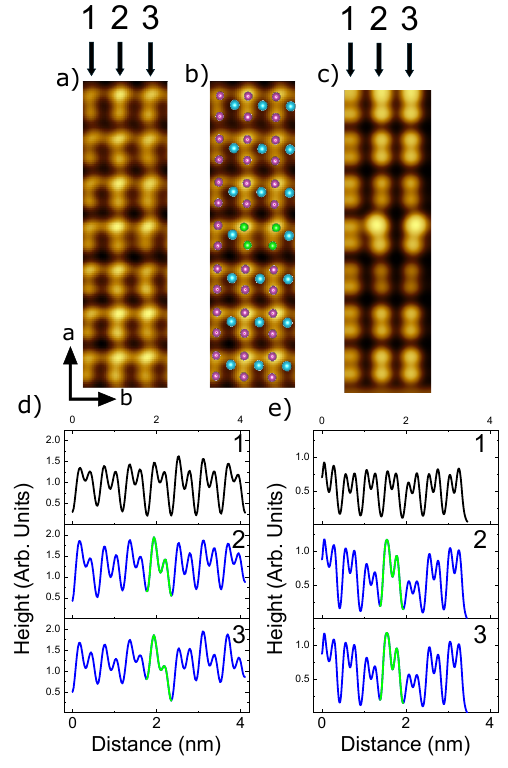}
\caption{EBD STM image and simulation. a) STM-acquired image (4.06 nm x 1.17 nm) zoomed in on an EBD (I = 50 pA, V$_{Sample}$ = -50 mV). b) Calculated relaxed structure with single Zr vacancy overlaid on the STM-acquired image from a). The ions shown in green are the four Te ions surrounding a Zr ion beneath and are color coded to help associate features in the linecuts of d/e with ions in the topographies. c) -50 mV simulated topography. d) Three \textit{a}-axis linecuts through STM image with linecuts 2 and 3 through the defect region. e) Corresponding three \textit{a}-axis linecuts through the simulated image showing general agreement with those through the STM image.}
\label{fig:Fig8}
\end{figure}

We next explore the possibility that the subsurface defect originates from a Zr vacancy. Figure \ref{fig:Fig8} shows a) high-resolution STM image of an EBD, b) an overlay of the relaxed structure for a Zr vacancy, highlighting correspondence between the locations of the relaxed atomic positions and ions in the STM image, and c) the resulting simulated STM topography. Here we emphasize the use of a 7x3x1 supercell in our calculations which, while computationally intensive, are not computationally prohibitive as were larger supercells attempted. As a result, while the essential details of extended defects are captured in our calculations, not all components and features of defects can be detailed.

\begin{figure}[t]
\includegraphics[clip=true,width=\columnwidth]{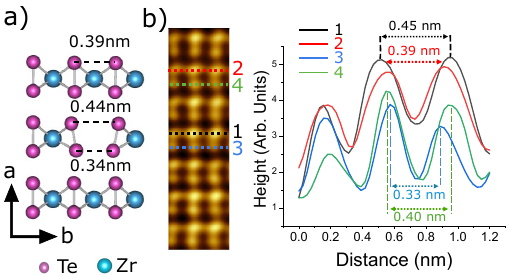}
\caption{a) Relaxed crystal structure of ZrTe$_3$ with subsurface Zr vacancy. Whereas the surface Te-Te spacing along the \textit{b}-axis is 0.39 nm away from the defect center, the Te ions nearest to the Zr vacancy center show an increase in the Te(2)-Te(2) spacing (0.44 nm) and a decrease in the Te(3)-Te(3) spacing (0.34 nm) b) Linecuts through the STM-acquired topography shows a very similar distortion of the four surface Te ions surrounding the defect center. Linecut 1 is through Te(2) ions directly above the center location and shows the Te(2) ions shift outward in comparison to the Te(2) ions of linecut 2, which are away from the center of the defect. Linecut 3 through Te(3) directly below the center location shows the Te(3) ions shift slightly inward in comparison to the Te(3) ions of linecut 4, which are away from the center of the defect.}
\label{fig:Fig9}
\end{figure}

First, evident from the Zr vacancy simulated image and associated linecuts is a raised or bright region spanning two neighboring atomic rows within the defect as compared to Te ions outside of the defected area, in agreement with the EBDs observed. Next, directly neighboring the defect location, there is a slight depression extending a few unit cells, in agreement with what is observed in the defect STM linecuts of Figure \ref{fig:Fig6}. Finally, in the simulated Zr vacancy image, there is an evident \textit{b}-axis structural distortion of the four Te ions closest to the vacancy location which is quantified in the calculated relaxed crystal structure (Figure \ref{fig:Fig9}). Such a distortion is expected as Zr ions are directly bonded to four surface Te ions. Linecuts through the STM image of an EBD along the \textit{b}-axis shows similar \textit{b}-axis distortions in the four surface ions surrounding the center of the defect (Figure \ref{fig:Fig9}). Taken together, we identify the EBDs as subsurface Zr vacancies.

We examined the possibility of subsurface Te vacancies as candidates for EDDs and EBDs but have eliminated them as possibilities.\cite{SupMat} Our DFT calculations indicate that subsurface Te vacancies result in a localized but minor surface structural distortion of four Te ions which are more challenging to detect in STM images as they lack an extended nature.\newline

\begin{figure*}[t]
\includegraphics[clip=true,width=\textwidth]{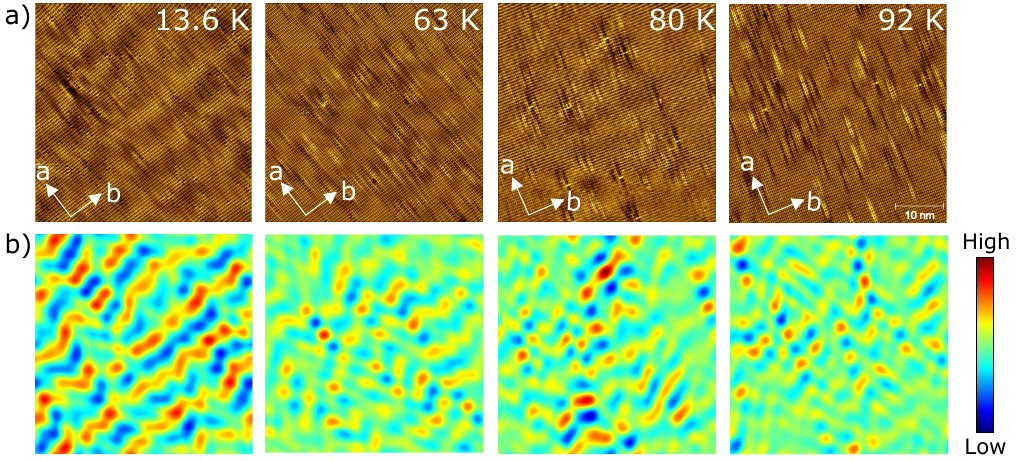}
\caption{Temperature evolution of the CDW. a) 45 nm x 45 nm topographic images taken below T$_{CDW}$ (13.6 and 63 K) as well as above T$_{CDW}$ (80 and 92 K). I = 50 pA, V$_{Sample}$ = -50 mV. b) Topographic images from a) Fourier filtered to extract only the CDW-associated signal.}
\label{fig:Fig10}
\end{figure*}

\begin{figure}[b]
\includegraphics[clip=true,width=\columnwidth]{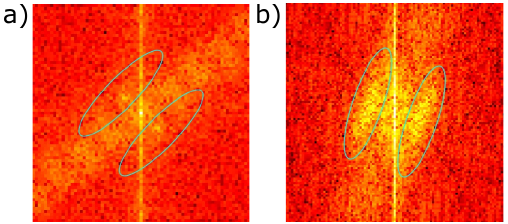}
\caption{FFTs at 9 K and 58 K. a) FFT of a 70 nm x 57 nm topography acquired at 9 K zoomed to the region containing the CDW signal. The CDW periodicity can be well characterized by a single wavevector of \textit{q}$_{CDW}$ =  0.07\textit{a}*. b) FFT of an even larger topography, 78 nm x 100 nm, acquired at 58 K showing a much more diffuse CDW signal (encircled by an ellipse) instead of a single distinct wavevector, consistent with the observed real space spatial variations of Figure \ref{fig:Fig10}. Note that both a) and b) are FFTs of topographies acquired below the bulk T$_{CDW}$ $\sim$ 72 K. The ellipses in both a) and b) encircle the same wavevector regions of the FFTs.}
\label{fig:Fig11}
\end{figure}

\noindent{\textit{CDW Temperature Dependence:}} \newline
We have conducted STM measurements from well below T$_{CDW}$ $\sim$ 72 K to above. Figure \ref{fig:Fig10}a shows topographic images acquired at 13.6 K, 63 K, 80 K and 92 K. Notably, above T$_{CDW}$, the CDW is still present, though there are localized regions in which the CDW appears absent. This is particularly obvious in the extended image of Figure \ref{fig:Fig3} acquired at 92 K where, by eye, the CDW state is preferentially localized by defects. Next, with temperature, spatial variations in the CDW stripes across the topography become prevalent even at temperatures below T$_{CDW}$, as has been observed previously \cite{Liu2021}. To highlight the variations in the CDW with temperature, we Fourier filter the topographic images to include only the CDW signal (Figure \ref{fig:Fig10}b). Associated with these spatial variations with temperature, the sharp FFT peaks associated with q$_{CDW}$ at low temperatures become more diffuse with temperature (Figure \ref{fig:Fig11}).

\begin{figure*}[!t]
\includegraphics[clip=true,width=\textwidth]{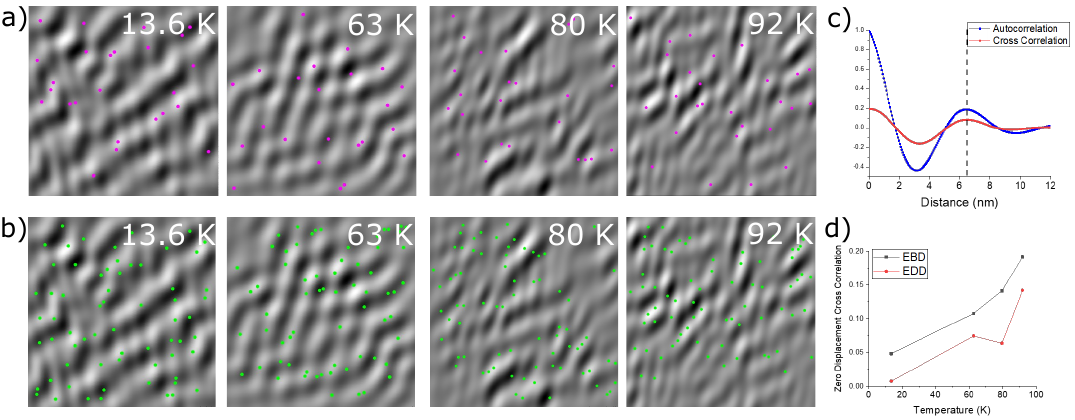}
\caption{a), b) Using template matching, locations of EBDs and EDDs respectively overlaid on filtered CDWs from topographies of Figure 10. c) Comparison of linecuts through the 92 K filtered CDW signal autocorrelation and the EBD defect-CDW cross-correlation in the direction of the CDW periodicity. Peaks in the cross-correlation occur at zero-displacement and coincide with the $\sim$6.5 nm CDW periodicity indicating a coupling of the CDW peaks to defect location. d) Zero-displacement cross-correlation values for defect locations (EBD and EDD) with the CDW signal with temperature. An overall increasing trend with temperature is seen for both types of defects.}
\label{fig:Fig12}
\end{figure*}

Next, we explore the possible connection of CDW variations to the observed EDDs and EBDs. While isolating the CDW signal in topographic images is straightforward using Fourier filtering, a systematic identification of all of the EDDs and EBDs across a region is more challenging given their aperiodic nature. To identify each type of defect, we use two methods which give similar results: 1) Fourier filtering of defect-related signals and 2) template matching.\cite{SupMat} Figure \ref{fig:Fig12} shows the locations of the identified EDDs and EBDs in the topographic images of Figure \ref{fig:Fig10}, overlaid on the Fourier filtered CDW state.

Having identified the EDD and EBD locations in our images, we are able to calculate the density of defects in our samples. For an EDD, originating from a Te surface vacancy, we find a density of 0.4\%. For an EBD, originating from a Zr vacancy, we find a density of 0.3\%. In each case, the defect density we calculate is the percentage of defects (EDDs or EBDs) as compared to the related total number of the specific type of ions (Te for EDDs or Zr for EBDs) which should present within a given imaged region. We note that within a given image, there are more than double the number of EDDs as compared to EBDs. However, given the crystal structure, there are half the number of Zr ions as compared to surface Te ions in a given region meaning that, were Zr and Te surface vacancies to occur with the same probability, we would expect EDDs to occur at exactly double the number of EBDs. In our measurements we find a similar, but smaller percentage of Zr vacancies than percentage of Te surface vacancies. Indeed, our DFT calculations indicate that the energy of formation of Zr vacancies is 100 Ry which is more than that of 29 Ry for a Te surface vacancy. In either case, the 0.3\% and 0.4\% identified vacancies are well below the 5\% Hf doping of our sample. This further supports that neither EDDs nor EBDs are due to Hf dopants.

Finally, to explore a possible connection between the CDW state and the defects in a quantifiable manner, we calculate the cross-correlation between the defect locations and the CDW signal. For the 92 K data (Figure \ref{fig:Fig12}), as an example, we show a comparison of the CDW signal autocorrelation and the EBD defect-CDW cross-correlation. This comparison illustrates a pinning effect of the CDW crests to the defect locations, a cross-correlation maximum found at zero displacement. Across the temperature range of this study, the zero-displacement cross-correlation significantly increases with temperature for both EDDs and EBDs with the CDW signal (Figure \ref{fig:Fig12}). In short, with temperature, CDW maxima are increasingly preferentially found at defect sites.

Further understanding this increasing correlation of defect locations with CDW maxima with temperature, we believe, is nuanced. Below T$_{CDW}$, where the CDW state is ubiquitous throughout the sample, an increase in correlation points toward an increasing pinning strength with temperature. This suggests an evolving competition between the energy considerations associated with elastic energy cost of CDW deformations and energy gained by defect pinning.\cite{Fukuyama1978, Ravy2006, Feinberg1988} Above T$_{CDW}$, where the CDW state has more spatial variation, appearing in more localized patches, an increasing correlation with temperature could point toward further increasing pinning strength with temperature or it could point to defects as nucleation sites for the CDW as has been found in other CDW materials \cite{Arguello2014}.

\subsection{Methods}
Temperature-dependent STM measurements were carried out using an RHK PanScan Freedom STM operating at $\sim$$10^{-9}$ Torr. The STM tip is mechanically cut Pt-Ir (80\%-20\%) which was cleaned and sharpened via electron bombardment. Details of the growth of the high quality Zr$_{0.95}$Hf$_{0.05}$Te$_3$ single crystals studied are found in \cite{Liu2022}. Samples were cleaved at $\sim$$10^{-8}$ Torr.

\subsection{Acknowledgments}
Work at Clark University was supported by the National Science Foundation under Grant No. DMR-1904918. Work at BNL was supported by the Office of Basic Energy Sciences, Materials Sciences and Engineering Division, U.S. Department of Energy (DOE) under Contract No. DE-SC0012704. This work was also supported by an Amrita Chancellor’s Fellowship from Amrita Vishwa Vidyapeetham.


\bibliography{ZrTe3}  

\end{document}